\documentclass[aps,prl,reprint,onecolumn,nofootinbib,groupedaddress]{revtex4-2}

\usepackage{amssymb}
\usepackage[english]{babel}
\usepackage{float}
\usepackage[letterpaper,top=2cm,bottom=2cm,left=3cm,right=3cm,marginparwidth=1.75cm]{geometry}
\usepackage{amsmath}  
\usepackage{amsfonts} 
\usepackage{graphicx} 
\usepackage[colorlinks=true, allcolors=blue]{hyperref}
\usepackage{xcolor}

\begin{document}
\title{Magnetic recoil interferometer in a uniform gravitational field. Comment on ``Observation of the quantum equivalence principle for matter-waves''}

\author{Peter Asenbaum$^{1}$, Chris Overstreet$^{2}$}

\address{$^1$ Institute for Quantum Optics and Quantum Information (IQOQI) Vienna, Austrian Academy of Sciences, Boltzmanngasse 3, 1090 Vienna, Austria.}

\address{$^2$ Department of Physics \& Astronomy, Johns Hopkins University, Baltimore, Maryland 21218, USA}

\begin{abstract}
The calculation of the phase shift of a matter-wave interferometer looks different for different coordinate choices or the addition of a uniform gravitational field, but the final result of the observable phase shift does not depend on the coordinate choices or the uniform gravitational field.
We find that the phase shift observed in ``Observation of the quantum equivalence principle for matter-waves'' is a magnetic recoil phase shift, similar to the recoil phase shift in light-pulse interferometers. Recoil phase shifts are inversely proportional to the inertial mass and do not depend on the gravitational mass. Our analysis highlights the difference between magnetic and gravitational acceleration in quantum mechanics.   
\end{abstract}
\maketitle

\textit{\textbf{Introduction.}} In Ref. \cite{dobkowski2025}, the authors measure a non-vanishing phase shift $\phi$ in a matter-wave interferometer and claim its dependence on a uniform gravitational field $g$ via $\phi = -m g^2 T^3 / 3\hbar$.
In this work, we distinguish between uniform gravitational and magnetic fields and show that the observed phase shift actually depends on the value of an applied magnetic gradient rather than the gravitational field. The equivalence principle, which states that gravitational effects cannot be measured locally and that uniform gravitational fields are unobservable \cite{DiCasola2015,Asenbaum2024}, still holds.

\textit{\textbf{Electromagnetic origin of the phase shift.}} In the experiment described in \cite{dobkowski2025}, an atom interferometer interacts with the magnetic field sourced by a current-carrying chip.  One interferometer arm, the ballistic wave packet, is in a magnetically insensitive state. The other interferometer arm, the reference wave packet, is in a magnetically sensitive state. The gradient of the magnetic field sourced by the chip keeps the reference wave packet at an approximately constant distance from the chip.

We calculate the phase shift of this interferometer in a uniform gravitational field $g_0$, without assuming the equality of the inertial mass $m_i$ and the gravitational mass $m_g$. For simplicity, we assume that the interferometer is perfectly closed and ignore phase shifts from the motion of the reference wave packet with respect to the chip,  i.e., inertial phase shifts. A similar analysis has been performed for clocks and symmetric matter-wave interferometers \cite{Asenbaum2024}. 
  
\begin{figure}[H]
\centering
\includegraphics[width=0.85\linewidth]{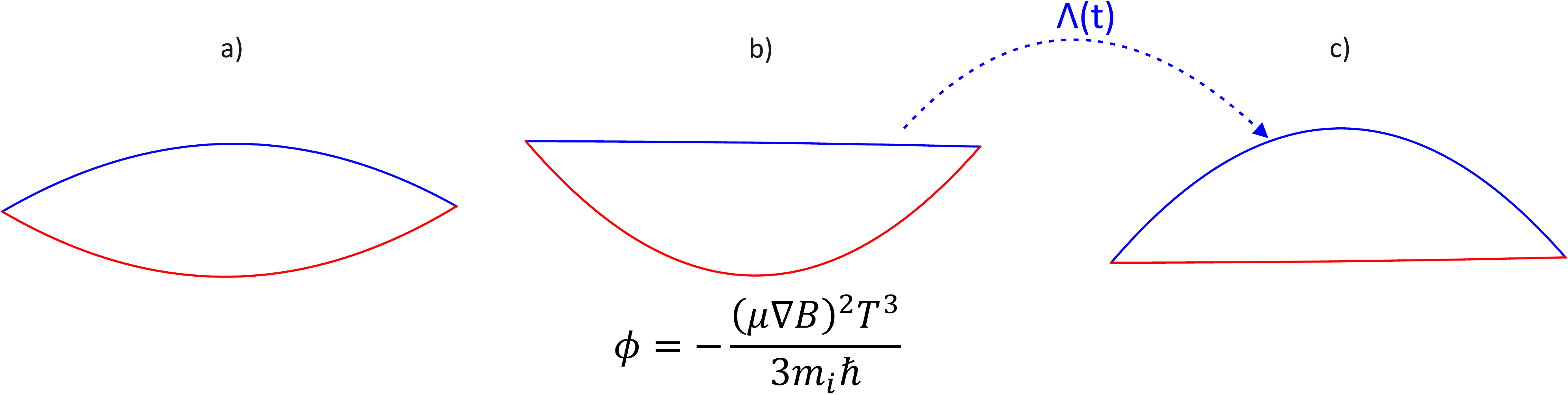}
\caption{Arm trajectories of the interferometer in \cite{dobkowski2025} for different values of the gravitational field $g_0$. Details of the phase shift calculation depend on the reference frame, but the phase shift is always the same and does not depend on $g_0$. \label{fig:1}}
\end{figure}

The reference wave packet (Fig. 1, red arm) is constantly accelerated by the magnetic force ${F_\text{mag}=\mu\,~\nabla B}$. The ratio of force and inertial mass $m_i$ gives the acceleration $F_\text{mag}/m_i$. The ballistic wave packet (Fig. 1, blue arm) is kicked upwards by a  short magnetic field pulse $B_{pulse}$, which leads to an initial velocity of $v_0=\mu \nabla B_{pulse}/m_i \cdot \tau$. The pulse strength is tuned so that $\nabla B_{pulse}\; \tau \approx\nabla B\; T$, which closes the interferometer after the full interferometer time $2T$. Keeping the initial velocity $v_0$ as an independent parameter would result in multiple magnetic terms with different time dependencies. To stay closer to the analysis in \cite{dobkowski2025}, we fix the two magnetic parameters to each other and set $v_0= F_{mag}/m_i \cdot T$.

In the presence of a uniform gravitational field (Fig. 1a or 1c), both arms have gravitational potential energy $m_g\, g_0z$ and accelerate at a rate $-m_g/m_i\, g_0$. In the coordinate system that corresponds to the Newtonian frame of Ref. \cite{dobkowski2025} (Fig. 1c), the ballistic wave packet has initial velocity $v_0$, and the reference wave packet is initially at rest.  The action of the ballistic wave packet is given by

\begin{align}
S^{(g)}_\text{ballistic} = \int_{0}^{2T} \left[ \frac{m_i}{2} \left( v_0 - \frac{m_g}{m_i} g_0 t\right)^2 -m_g g_0 \left( v_0 t - \frac{m_g g_0}{2 m_i}t^2\right) \right] dt \\
=\frac{F_{mag}^2}{m_i} T^3 - 4 F_{mag} \frac{m_g}{m_i} g_0 T^3 + \frac{8 m_g^2}{3 m_i}\, g_0^2  T^3
\end{align}
and the action of the reference wave packet is given by 
\begin{equation}
S^{(g)}_\text{reference} = \int_{0}^{2T} \left[ \frac{m_i}{2} \left(\frac{F_{mag}}{m_i}t - \frac{m_g}{m_i} g_0 t\right)^2 -m_g g_0 \left(\frac{F_{mag}}{2m_i} t^2 - \frac{m_g g_0}{2m_i} t^2\right) \right] dt.
\end{equation}
The phase shift is given by 
\begin{equation}
\phi = \frac{1}{\hbar} \left[ S^{(g)}_\text{ballistic} - S^{(g)}_\text{reference} \right] = -\frac{F_\text{mag}^2 T^3}{3 m_i\hbar} = -\frac{(\mu \nabla B )^2 T^3}{3 m_i\hbar}.
\end{equation}

Instead of calculating the action of the ballistic wave packet, one can start from the ``Einsteinian'' frame (Fig. 1b), in which the ballistic wave packet rests at $z_E$ and the gravitational field vanishes. In this frame, the action of the ballistic wave packet is trivially zero. The Einsteinian and Newtonian coordinates are related via  $z_E=z_N - (F_{mag}/m_i) T t+ (m_g\, g_0/2 m_i) t^2$, which leads to a gauge phase $\Lambda(t)$ between the frames.  At the end of the interferometer, the gauge phase 
\begin{equation}
\Lambda(2T) = \frac{1}{\hbar} \left[\frac{F_{mag}^2}{m_i} T^3 - 4 F_{mag} \frac{m_g}{m_i} g_0 T^3 + \frac{8m_g^2}{3 m_i}\, g_0^2 T^3 \right]
\end{equation}
is proportional to the action of the ballistic wave packet in the Newtonian frame, $\Lambda(2T)= (1/\hbar) S^{(g)}_\text{ballistic}$.  
This gauge phase can be combined with the action of the reference wave packet $S^{(g)}_\text{reference}$ to compute the observed interferometer phase, leading to the same result as before.  

Regardless of how it is calculated, the phase shift observed in the experiment is given by 

\begin{equation}
\boxed{\phi={-\frac{(\mu \nabla B )^2 T^3}{3 m_i\hbar}}.}
\label{recoil}
\end{equation}
Crucially, the phase shift depends on the magnetic gradient and the inertial mass only, not on the uniform gravitational field $g_0$.
It has the form of a recoil phase shift, which scales as momentum difference squared over the inertial mass. The term \textit{recoil} indicates that the phase shift arises from kinetic energy differences \cite{Bongs2006}. The gravitational acceleration is common between the arms and, whatever its strength may be, creates no phase shift. The magnetic force is differential between the arms and causes the magnetic recoil phase shift in Eq. \ref{recoil}.

\textit{\textbf{Equivalence principle tests.}}  Since the Newtonian and Einsteinian reference frames provide the identical prediction for the phase shift of the interferometer in Ref. \cite{dobkowski2025}, comparing the observed phase shift to this prediction does not test the equivalence principle.   
\textit{Inertial phase shifts} measure the motion of the matter waves with respect to a ruler, e.g., the phase of a laser beam or a magnetic field. So far, we have neglected residual motion between the chip sourcing the magnetic fields and the reference wave packet. For a residual acceleration $a_{res}$, the inertial phase shift is proportional to $a_{res} F_{mag}T^3/\hbar$. Inertial phase shifts depend on the inertial to gravitational mass ratios of the atoms and the reference system (e.g. a different atomic species or a falling corner cube). 
The interferometer in Ref. \cite{dobkowski2025}  can be understood as an equivalence principle test between the atoms and the chip, when characterizing the precision of the magnetic phase shifts. The relationship between an inertial phase shift and a measurement of a non-uniform gravitational field is discussed in the appendix of \cite{Asenbaum2024}.

\textit{\textbf{Concluding remarks.}} The equality of inertial and gravitational mass has been confirmed to high precision. Therefore, the acceleration caused by a magnetic force and the acceleration caused by a uniform gravitational field are very different.  The magnetic force on a test particle depends on its properties (e.g. its magnetic moment) and is observable, but uniform gravitational fields act identically on all physical systems and are not observable \cite{Asenbaum2024}.

The standard definition of the equivalence principle states that gravity is not observable locally \cite{DiCasola2015}.  This definition allows the equivalence principle to be tested to high precision in local experiments.  Formulations of the equivalence principle that compare physics in a uniform gravitational field with physics in a faraway ``gravity-free space'' are problematic, as it is not clear how to test them.

\end{document}